# Hexagonal $Sr_{0.6}Ba_{0.4}MnO_3$ : Spin and Dipole Coupling via Local Structure

Ritu Rawat[a*], R.J. Choudhary[a], A.M. Awasthi[a*], Archna Sagdeo[b], A.K. Sinha[b], Rajamani Raghunathan[a], V.G. Sathe[a], and D.M. Phase[a]

[a]UGC DAE Consortium for Scientific Research, Indore- 452 001, India

[b]Indus Synchrotrons Utilization Division, Raja Ramanna Centre for Advanced Technology, Indore- 452 013, India

**Abstract**

Hexagonal $Sr_{0.6}Ba_{0.4}MnO_3$ (SBMO) follows $P6_3/mmc$ symmetry where $MnO_6$ octahedras are both face ($Mn_2O_9$ bioctahedra) and corner shared via oxygen anion. It undergoes ferroelectric (FE) and antiferromagnetic (AFM) orderings close to the room temperature. Magnetic properties appear to be governed by intricate exchange interactions among $Mn^{4+}$ ions within and in adjacent $Mn_2O_9$ bioctahedra, contingent upon the local structural changes. Calculations based on our model spin-Hamiltonian reveal that the dominant linear AFM fluctuations between the $Mn^{4+}$ ions of two oxygen-linked bi-octahedra result in short range correlations, manifest as a smooth drop in magnetization below 325K. Competition between spin-exchange and local-strain is reckoned as responsible for the atypical magneto-electricity obtained near the room temperature.

*Corresponding Authors: ritu@csr.res.in, amawasthi@csr.res.in

**Introduction:**

In recent years, in some magneto-electric materials the prerequisite of $d^0$-ness for ferroelectricity (FE) is circumvented [1-3]. Rather, the FE-origin in them is attributed to particulate displacements of atoms carrying a net spin, responsible for the inversion-symmetry breaking. The resulting softening of the appropriate phononic modes is coupled with the magnetic ordering in the material [4, 5], thus realizing a novel mechanism for magneto-electricity. $AMnO_3$ (AMO, A =Ca and Sr) based materials are a revolution for magneto-electric multiferroics, as ferroelectricity and magnetism are derived by the same $Mn^{4+}$ cation. Recently, Bhattacharjee et. al. showed that with increase in the A cation size in $AMnO_3$ (AMO, A =Ca and Sr), FE distortion is strongly favored [6]. Also, since the $Mn^{4+}$ d-orbitals have degenerate $t_{2g}^3$ configuration in AMO, they contribute to the magnetic character whereas, the empty $e_g^0$ acts analogous to the $d^0$ configuration, and participates in the ferroelectric character [7]. Sakai et al. reported that ferroelectricity in cubic $Sr_{1-x}Ba_xMnO_3$ is induced due to the displacement of $Mn^{4+}$ ion and modulation in O-Mn-O bond angle, because of the different ionic radii of Ba and Sr. For half-doped cubic $Sr_{1/2}Ba_{1/2}MnO_3$, though the ferroelectric transition is observed at ~400K, magneto-electricity (ME) occurs at and below the Néel temperature $T_N$ ~185K [3].

Well-studied robust multiferroics with long-range ordering of both their spin and dipolar degrees of freedom (having separated/Type-I or concurrent/Type-II transitions) exhibit dielectric anomalies sharply confined onto their magnetic transition temperature [8]. Lately, a new class of multiferroics has appeared with either one or both these degrees of freedom developing only short-range correlations [9]. Examples of this so-called Type-III multiferroics include e.g., multiglasses (Sr,Mn)$TiO_3$ [10], antiferromagnetic quantum paraelectric glass $SrCu_3Ti_4O_{12}$ [11], multiglass $La_2NiMnO_6$ [12] etc. In the present study, we unravel the manifestation of ferroelectric and antiferromagnetic orderings in hexagonal $Sr_{0.6}Ba_{0.4}MnO_3$ (SBMO) within a narrow ambient range close to the room temperature. Coupling of magnetic and electrical properties via structural and vibrational degrees of freedom is experimentally established.

**Experimental Details:**

Polycrystalline hexagonal $Sr_{0.6}Ba_{0.4}MnO_3$ (SBMO) was prepared using solid state reaction method. Purity of the sample is confirmed using X-ray diffraction measurements and Rietvield and Le Bail fitting. X-ray diffraction measurements were performed using table top D2 phaser Bruker X-Ray diffractometer. An energy dispersive analysis of X-ray (EDAX) unit attached to FEI NOVA Nano SEM 450 was used to record EDAX pattern. Near edge X-ray absorption spectra was recorded at SXAS beamline BL-1, Indus-2, RRCAT, Indore. Raman measurements were performed using a HR800 Jobin-Yvon spectrometer. He-Ne laser with a wavelength of 632.8 nm was used to record Raman spectra. Magnetization measurement was carried out using 7T SQUID VSM (Quantum Design, USA). Dielectric measurements over 10kHz to 1MHz were performed in the temperature range of 250K to 410K in cooling cycle, using Alpha-A high performance frequency analyzer (Novo Control). Specific heat $C_p(T)$ was obtained at 5°C/min warm up rate from STAR$^E$ DSC-1 (Differential Scanning Calorimeter, Mettler-Toledo). Temperature dependent XRD measurements were performed using Synchrotron source at RRCAT Indus-2, BL-12. Field dependent isothermal (room temperature) magneto-dielectric measurements at 1MHz were performed using Alpha-A high-performance frequency analyzer (Novo Control) and a 9Tesla Integra cryostat/magnet (Oxford NanoSystems).

**Results and Discussion:**

XRD analysis of the SBMO sample with Rietvield refinement and Le Bail fitting performed using Full Prof software confirms its single phase nature with hexagonal P6$_3$/mmc symmetry (shown in Fig. 1(a)) [13]. The obtained lattice parameters and the Wyckoff positions are shown in Table 1. We have also carried out near edge X-ray absorption spectra (XANES) studies at Mn-L edge of SBMO, as shown in Fig. 1(b), which confirm the 4+ state of Mn, consistent with the stoichiometric oxygen content in the system. EDAX measurement also confirmed the stoichiometry of SBMO (shown in Fig. 1(c)). Magnetization measurements reveal magnetic anomaly in the form of a broad maximum at temperature $T_M$ ~325K, as shown in Fig. 2(a). Linear *M-H* behavior at 300K (inset of Fig. 2(a)) evidences its antiferromagnetic character. In SBMO, two face-shared MnO$_6$ octahedra stacked along the *c*-axis form bi-octahderal Mn$_2$O$_9$. This gives rise to two Mn-O-Mn superexchange interactions; a linear/180° Mn-O1-Mn ($J_1$) and a non-linear/78.36° Mn-O-Mn ($J_2$) one (bond angle estimated from the analysis of the XRD pattern, Inset Fig. 1(b)). It is inferred that the antiferromagnetic interaction between Mn$^{4+}$ ions of linear Mn-O1-Mn bonds leads to short range correlations (SRC) and, together with the weak interactions between Mn$^{4+}$ ions of non-linear Mn-O-Mn bonds, give rise to a broad feature at 325K, as discussed later.

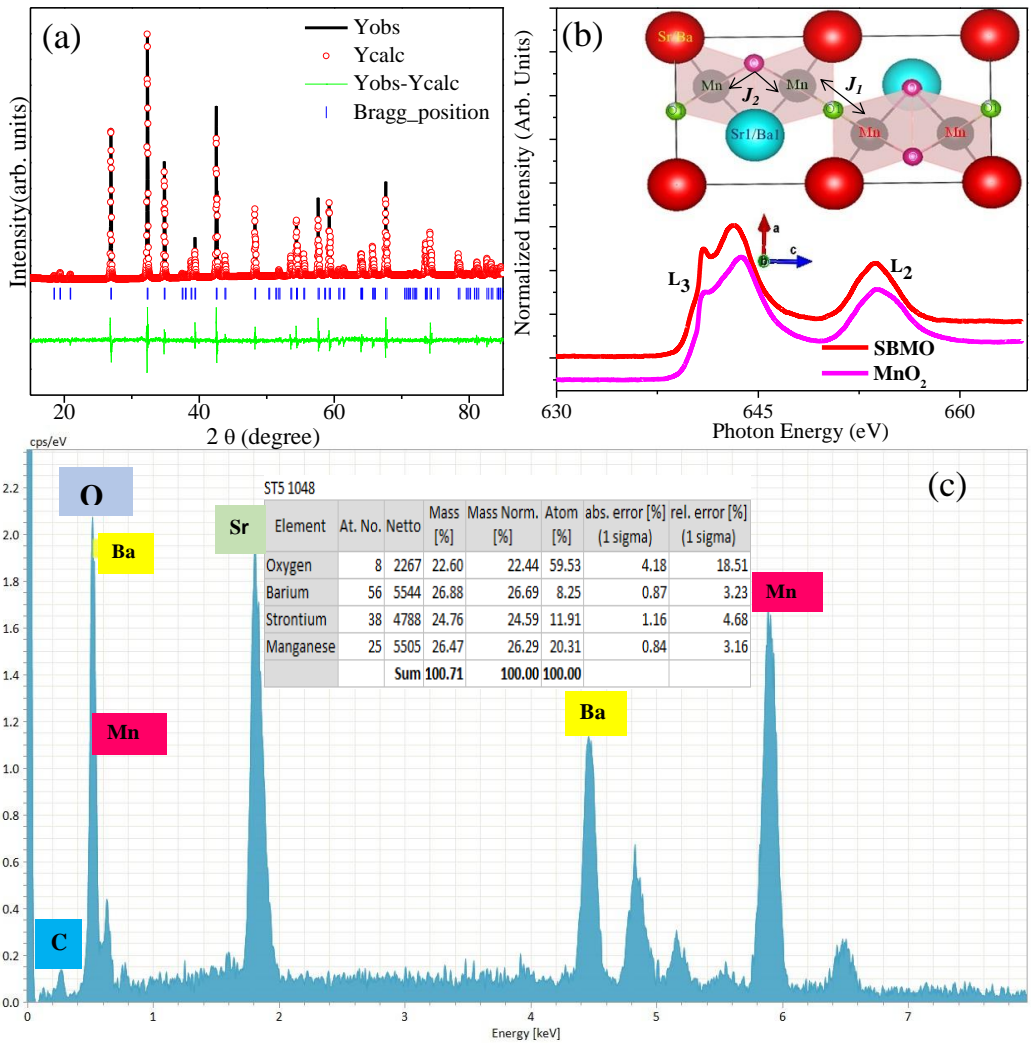

**Fig.**1 (a) XRD pattern using lab source fitted with Rietvield refinement and Le-bail fitting. (b) Mn L-edge XANES of SBMO and reference $MnO_2$ sample. Inset shows arrangement of atoms for $P6_3/mmc$ symmetry in *zx*-plane, drawn using the Vesta software (Note: two sites of oxygen-- O and O1, & interactions $J_1$ and $J_2$). (c) EDAX spectra of SBMO sample with elemental composition.

Table I: The obtained lattice parameters and the Wyckoff positions of SBMO.

| | T=300K   $P6_3/mmc$   a=b=5.537Å c=9.149Å | | | |
|---|---|---|---|---|
| Atom | Wyckoff | x | y | z |
| Sr1 | 2a | 0 | 0 | 0 |
| Sr2 | 2c | 1/3 | 2/3 | 1/4 |
| Ba1 | 2a | 0 | 0 | 0 |
| Ba2 | 2c | 1/3 | 2/3 | 1/4 |
| Mn | 4f | 1/3 | 2/3 | z |
| O1 | 6g | 1/2 | 0 | 0 |
| O | 6h | -x | x | 3/4 |
| | z=0.61325 | x= 0.82151 | | |

Dielectric constant versus temperature features nearly concurrent peaks at ≈355K for various frequencies (Fig. 2(b)). As we do not get stable and robust *P-E* hysteresis loops across this temperature, we understand this as signaling dynamic dipole correlations, indicative of incipient ferroelectricity emergent below "$T_C$"~355K, as also observed in the case of mesoscale-sized particles of SmFeO3 [14].

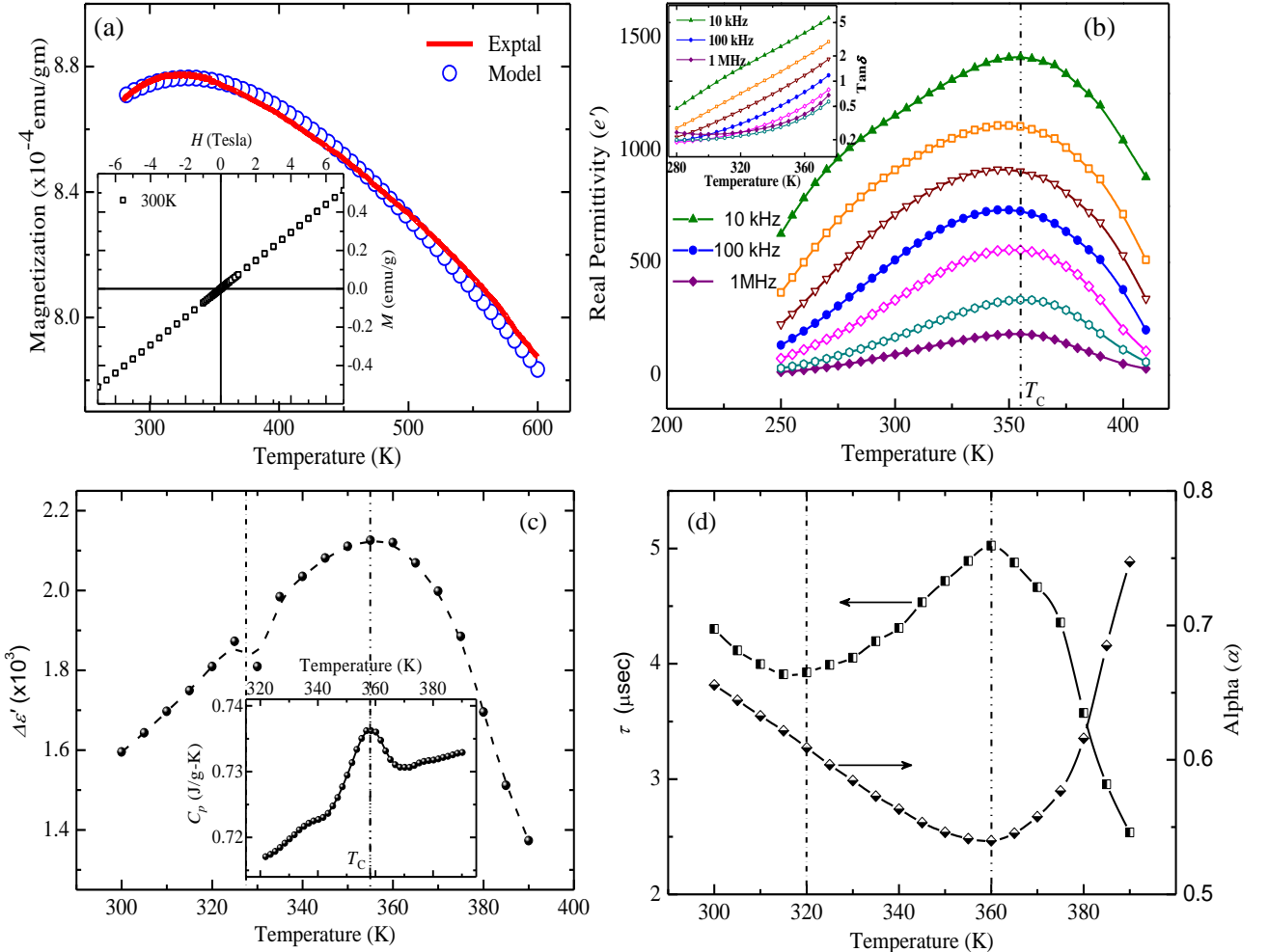

**Fig.2** (a) Magnetization vs. temperature fits (continuous lines) of SBMO. Inset shows the *M-H* behavior at 300K. (b) Dielectric constant $\varepsilon'(T)$ at selected high frequencies. Inset shows loss tangent across $T_M$ and $T_C$. (c) Dielectric strength ($\Delta\varepsilon'$) versus temperature. Inset shows the associated $C_p$-peak near the incipient FE-$T_C$. (d) Relaxation time $\tau(T)$ in the left pane and the spectral broadening index $\alpha(T)$ in the right pane.

The somewhat smeared peaks seen here in $\varepsilon'_\omega(T)$ are not rooted in generic dipolar relaxation, since the associated loss tangent values (Fig. 2(b), inset) are devoid of any otherwise-necessitated peaks/maxima. In a very recent review, Scott and Gardner [15] observed that in typical magneto-electric materials, somewhat 'imperfect/distorted FE signatures' may indicate incipient FE, which grows into robust/long-range ferroelectricity in their nanostructured/thin-film formations. The dielectric constant drop upon cooling is seen to slowdown some ~30K below $T_C$. Emergence of these relatively frequency-indifferent plateau at ~320K indicates spin-order driven/concurrent stabilization of the ferroelectric fluctuations— resulting in local static ferroelectricity, commensurate with the short-range magnetic ordering at $T_M$.

The above inferences are further confirmed by fitting the empirical Cole-Cole function onto the dielectric spectra spectra $\varepsilon^*(\omega) = \Delta\varepsilon/[1+(i\omega\tau)^{\alpha}]$ over 300-400K temperature range. For a number of manganites, Shrettle et al. [16] established that upon magnetic ordering, a robust FE transition is marked by sharp maxima in their relaxation time ($\tau$) and dielectric strength ($\Delta\varepsilon$), and minima in symmetrical broadening index ($\alpha$); all signaling a critical slowdown of the relaxations. Moreover, the development of only short-range dipole correlations registers a step (cf., 'negative peak') in the relaxation timescale. Since in our case here, only incipient (dynamic/scale-free) FE is indicated, and the magnetic ordering too happens to be short-ranged, the expectedly widened out peak feature about $T_C$ and step/negative-peak feature about $T_M$ are indeed obtained, as shown in Figs. 2(c, d); consistent with the rather low values of $\alpha$, which implies a very broad distribution of relaxation times. The two features continuously join (albeit well-resolved in $T$, as marked by the vertical lines), due to their considerable broadening. To the best of our knowledge, these are the maiden 'relaxational' signatures of both scale-free/dynamic and locally-static ferroelectric correlations. Intrinsic nature of the incipient FE is also manifest in specific heat ($C_p$-$T$, inset of Fig. 2(c)), depicting a wide peak at 357K near the designated '$T_C$'.

In order to elucidate the variation of local structure responsible for the above-asserted alterations in the 'ferroelectric' character, temperature dependent XRD patterns were recorded using Synchrotron source in the temperature range 300 to 350K. XRD patterns in the studied temperature range reveal that the crystal symmetry remains P6$_3$/mmc, thereby ruling out a robust (long range) ferroelectric ordering. In Fig. 3(a) we have plotted the difference in Mn-O and Mn-O1 bond lengths (Δ) versus temperature. At 350K, we observe inequivalent Mn-O and Mn-O1 bond lengths. Here, in MnO6 octahedra, bond length between Mn-O1 and Mn-O differs by ~0.05Å (inset of Fig. 3(a)). Therefore, inversion symmetry is locally broken in SBMO, which is the driving factor for the observed incipient and local FE-characters, although SBMO maintains overall centrosymmetric P63/mmc structure over the studied temperature range. It should be noted that though Rietveld analysis provides information related to an average long-range structure, yet at local or short range level, the structure can adopt non-centrosymmetric character, yielding inequivalent bond lengths [17-19]. It is also observed from Fig. 3(a) that Δ increases with decrease in temperature down to ~330K, reflecting the consolidation of local inversion symmetry breaking.

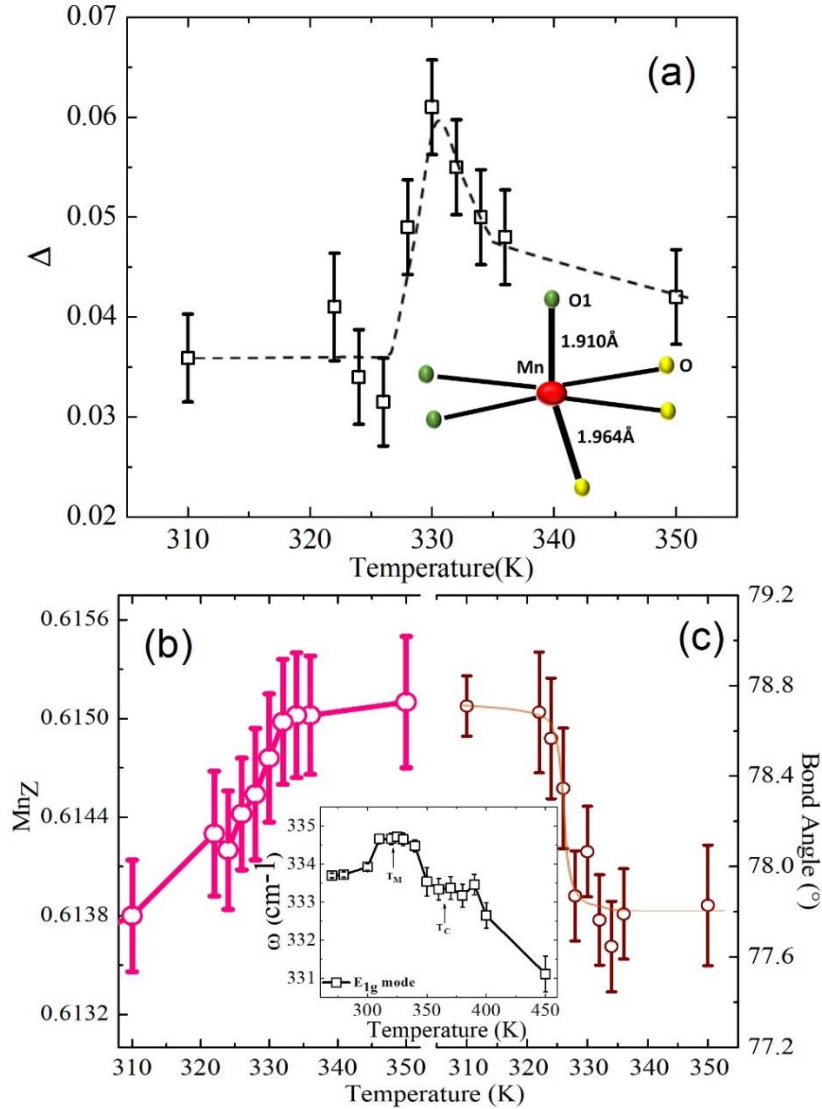

**Fig.3** (a) Difference between Mn-O1 and Mn-O bond lengths (Δ) plotted versus temperature (solid line is guide for the eyes); inset shows the $MnO_6$ octahedra, (b) shift in Wyckoff position of Mn along *z*-direction with respect to temperature, and (c) shows change in bond angle Mn-O-Mn versus temperature. Inset in (b) shows Raman shift of $E_{1g}$ mode plotted versus temperature.

From Figs. 3(b, c), which show the shift in Mn position along the $z$-direction and corresponding Mn-O-Mn bond angle respectively, it is evident that across $T_M$, antiferromagnetic exchange $J_1$ significantly $z$-shifts the Mn-atoms. The $z$-shift of Mn causes ambivalent variations in $\Delta$, with its implications on the local-stabilization of ferroelectric fluctuations, as discussed above. Similar such observation was also made by Singh et. al. in $0.9BiFeO_3$-$0.1BaTiO_3$ composite, in which, while the structure remains the same, change in bond-angle across the magnetic transition temperature and hence shifts in the position of atoms attribute to the origin of ferroelectricity [20].

Now we discuss the rationale for the observed magnetic behavior. In SBMO, an intricate magnetic behavior is anticipated. The unit cell (*u.c.*) of this 4H structure has two face-sharing $Mn_2O_9$ bi-octahedra; which in turn link together in a corner-sharing fashion through the common oxygen (inset Fig. 1(b)). The corner-shared linkage leads to a 180° antiferromagnetic superexchange (AF-SE) ($J_1$) in the network. Further, within the bi-octahedra, the short $Mn^{4+}$-$Mn^{4+}$ distances (~2.50 Å) and the ~80° $Mn^{4+}$-O-$Mn^{4+}$ triad through the face-shared oxygen atoms leading to direct- ($J_D$) and super-exchange ($J_S$) interactions respectively, giving rise to an effective magnetic exchange $J_2$. The actual strength and sign of the magnetic interactions $J_1$ and $J_2$ depends on the extent of the orbital overlap between the magnetic ion and the intervening oxygen [21]. Convolution of the microscopic exchange interactions at work over particulate temperature windows, effectively determines the magnetization status of the system therein.

The magnetic data can be modeled by considering the Hamiltonian,

$$\hat{H} = -J_1 \hat{S}_2 \cdot \hat{S}_3 - J_1 \hat{S}_4 \cdot \hat{S}_1 - J_2 \hat{S}_1 \cdot \hat{S}_2 - J_2 \hat{S}_3 \cdot \hat{S}_4 \quad (1)$$

where, $J_1$ ($J_2$) corresponds to the strength of inter (intra) -bioctahedral exchange interaction, $\hat{S}_i$'s are the spin operators, and the subscripts on the spin operators correspond to the indices of $Mn^{4+}$ sites in a *u.c*. The interaction between sites '4' and '1' imposes the periodic boundary condition in the *u.c*. Positive (negative) values of $J_i$ correspond to ferromagnetic (antiferromagnetic) interactions. Nalecz et al. [22] have shown for 4H-$SrMnO_3$ using electron paramagnetic resonance that due to the spin–orbit coupling and the crystal-field effects, Mn4+ quartet corresponding to spin 3/2 splits into two doublets: ±1/2 and ±3/2, with the doublet ±1/2 being lower in energy. These levels are shown to further split into non-degenerate levels under the application of magnetic field. In our spin-only model in the present study, the crystal-field effects and spin-orbit interactions are not explicitly taken into account. However, the spin population of Mn4+ ion in presence of an octahedral crystal-field is given by $t_{2g}^3$ yielding spin-3/2 at every Mn site and the same is considered in our calculation. Similarly, the spin-orbit interaction of the ions is included in the form of single-ion anisotropy term given by $DS_z^2$ as discussed below in equation 2.

The model Hamiltonian in eq.(1) conserves both total $\hat{S}^2$ and $\hat{S}_Z$ operators and hence it is possible to construct the Hamiltonian matrix (*H*) either in total-*S* or in total-$M_S$ basis. However, in this case the *H*-matrix is constructed in constant-$M_S$ basis, and then the eigenstates $E(S, M_S)$ of the spin model are numerically obtained by full diagonalization [23]. The expectation values of the $\hat{S}^2$ and $\hat{S}_Z$ operators for each eigenstate are obtained, from which the total spin *S* and $M_S$ of the state are deduced. Finally, the magnetization of the system as a function of temperature (*T*) at a chosen magnetic field ($H_Z$) is obtained from the relation,

$$M(H,T) = N_A g\mu_B \frac{\sum_S \sum_{M_S} M_S e^{-\frac{[E(S,M_S) - g\mu_B H_Z M_S + DM_S^2]}{k_B T}}}{\sum_S \sum_{M_S} e^{-\frac{[E(S,M_S) - g\mu_B H_Z M_S + DM_S^2]}{k_B T}}} \quad (2)$$

where, $N_A$ is the Avogadro number, $\mu_B$ is the Bohr magneton value, *g* is the Lande *g*-factor (taken to be 2.0), and magnetic field $H_z$ is set to 0.01 Tesla, as per the experiments and D the magnetic anisotropy. The magnetic data in the temperature range 300-600K is fitted by iterating over the values of $J_1$ and $J_2$ (Fig. 2(a)). Our best fit for the magnetization data in this temperature region yields $J_1 = -14.5$ meV, $J_2 = -1.3$ meV and D=6.3 meV. The computed exchange parameter values are comparable to those of 4H-$SrMnO_3$ and orthorhombic $LaMnO_3$ systems obtained using spin Hamiltonian [22, 24]. We further notice that the value of magnetization for the highest temperature reported is far less than the *spin-only* moments of the four uncorrelated spin-3/2 species per *u.c.* i.e., 3.3x$10^{-3}$ emu/gm; suggesting that even at 600K, the system has not transformed fully into the paramagnetic state. It should be noted that contrary to the previous theoretical studies on hexagonal $SrMnO_3$ [25], our experimental results on the crystal structure, Raman spectra, and magnetic measurements of SBMO, supported by our theoretical model, show that the broad hump occurring at $T_M$ =325K is a result of the AFM correlations between the $Mn^{4+}$ ions of two *adjacent* bi-octahedra. Neutron diffraction experiments as well as microscopic electronic models that include electron-phonon interactions ought to deepen our understanding of the complex magnetism in this system, which needs to be further explored.

The observed magneto-structural-vibrational coupling is further manifest in the Mn-displacement $E_{1g}$ Raman mode shift, as shown in the inset of Fig. 3(b). Across the incipient FE anomaly at $T_C$ and the magnetic ordering at $T_M$, the mode-position $\omega_s(T)$ deviates from its usual thermal anharmonic behavior. Since the buildup of ferroelectric correlations is expected to soften the phonon mode; therefore, due to its competition with the background anharmonic-rise, a fleeting 'stagnation' of the Raman shift is observed just below $T_C$. On the other hand, a net softening of the phonon mode observed below $T_M$ (overriding the anharmonic-rise in $\omega_s$) confirms the concurrent emergence of stable local ferroelectricity, upon the short-range magnetic ordering. These findings evidence that the entangled local-strain and spin-exchange interactions cause anomalies in the lattice vibrations across the $T_M$-values, confirming the spin-phonon coupling in SBMO [26], also observed in strained $EuTiO_3$, cubic $SrMnO_3$, and Ba-doped cubic $SrMnO_3$ multiferroics— through softening of their magnetic-ion related phonon mode— as a crucial signature for multiferroicity [4,5].

Magneto-dielectricity MD(%) is given by [27]

$$\frac{\varepsilon'(H) - \varepsilon'(0)}{\varepsilon'(0)} \times 100$$

We measured room temperature MD(*H*) isotherm under magnetic field up to ~7T at 1MHz, shown in Fig. 4, which is observed to be all positive. To discern if and how the (magneto) conductance influences our results, we have also plotted the magneto-loss [27] $\left\{ ML(\%) = \frac{\tan\delta(H) - \tan\delta(0)}{\tan\delta(0)} \times 100 \right\}$ at 300K in Fig. 4. Linear MD(*H*) and low-lying/flat ML(*H*) (solid lines) yielding |MD/ML|$_{RT}$ ~ +(10-13 dB) unambiguously ascertain robust and genuine magneto-dielectricity over ~2-5 Tesla field-window. All-negative ML$_{RT}$ plotted against all-positive MD$_{RT}$ in the inset (both implicitly varying vs. *H*) develops a co-regression with a negative slope above ~5T. Comparing with a model analysis by Catalan [28], this behavior is a signature that the observed high-field magneto-capacitance MD$_{RT}$(>5T) is traceable to the magneto-resistive character within the bulk. Thus, the room temperature ME character is rendered rather conspicuous at higher fields. We finally remark that the genuine ME as demonstrated adequately guarantee the moderate-field ME as originating from an essentially intrinsic dielectric response.

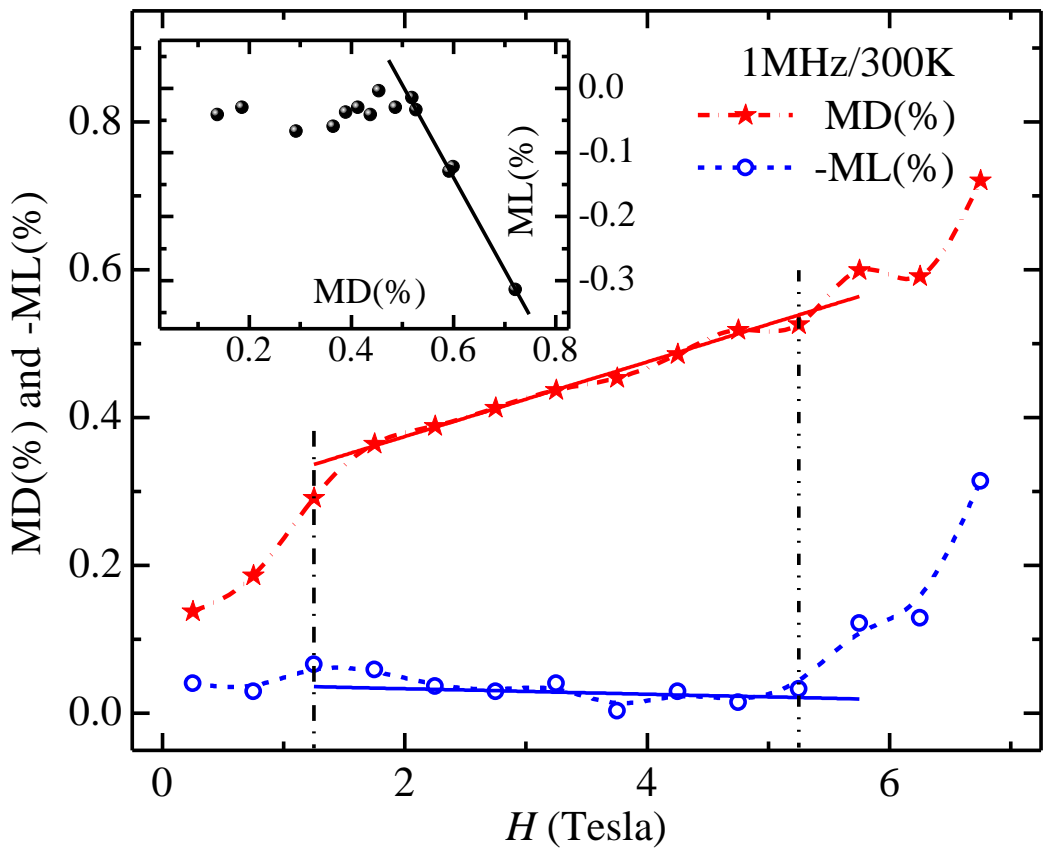

**Fig.4** Magneto-dielectricity (MD) and magneto-loss (ML) versus the applied field $H$ at 1MHz/300K. Inset reveals a changeover from the genuine magneto-electricity below ~5 Tesla to the circumstance dominated by the bulk magneto-resistance effects at the higher fields.

**Conclusion:**

Summarizing our work, it is established that the improper and short-range ferroelectricity manifest in $Sr_{0.6}Ba_{0.4}MnO_3$ arise from the structure-conserving/locally-unequal Mn-O bond lengths in the $MnO_6$ octahedra. The spin coupled Mn-displacement soft-phonon mode is responsible for magneto-electricity here. The broad magnetization-maximum and the concurrent anomalies in the micro-structure induce the change from incipient to stable short-range ferroelectricity. From our study, SBMO comes across as a rare material, where the FE and AFM orderings occur within a narrow temperature window close to the ambient. Multiferroics with competing spin-&-strain energetics thus offer a novel controllable route to practical ME-functionality. In this perspective it is vital to explore their nanostructures and oriented films, which in bulk form exhibit soft-mode FE ordering.

**Acknowledgements:** Authors are thankful to Dr. N.P. Lalla and Ms. Poonam Yadav of UGC-DAE CSR, Indore, for providing zero-field dielectric data. Authors are also thankful to Mr. Anupam Jana and Mr. Gyanendra Panchal for help in recording EDAX and XAS spectra respectively.

**References:**

[1] S. W. Cheong and M. Mostovoy, Multiferroics: a magnetic twist for ferroelectricity, *Nat. Mater.* **6** (2007)13.

[2] D. K. Pratt, J. W. Lynn, J. Mais, O. Chmaissem, D. E. Brown, S. Kolesnik, and B. Dabrowski, Neutron scattering studies of the ferroelectric distortion and spin dynamics in the type-1 multiferroic perovskite $Sr_{0.56}Ba_{0.44}MnO_3$, *Phys. Rev. B* **90** (2014) 140401(R).

[3] H. Sakai, J. Fujioka, T. Fukuda, D. Okuyama, D. Hashizume, F. Kagawa, H. Nakao, Y. Murakami, T. Arima, A. Q. R. Baron, Y. Taguchi, and Y. Tokura, Displacement-Type Ferroelectricity with Off-Center Magnetic Ions in Perovskite $Sr_{1-x}Ba_xMnO_3$, *Phys. Rev. Lett* **107** (2011) 137601.

[4] C. J. Fennie and K. M. Rabe, Magnetic and Electric Phase Control in Epitaxial $EuTiO_3$ from First Principles *Phys. Rev. Lett*. **97** (2006) 267602.

[5] S. Kamba, V. Goian, V. Skoromets, J. Hejtmˊanek, V. Bovtun, M. Kempa, F. Borodavka, P. Vanˇek,. A. A. Belik, J. H. Lee, O. Pacherovˊa, and K. M. Rabe, Strong spin-phonon coupling in infrared and Raman spectra of $SrMnO_3$, *Phys. Rev. B* **89** (2014) 064308.

[6] S. Bhattacharjee, E. Bousquet, and P. Ghosez, Engineering Multiferroism in $CaMnO_3$, *Phys. Rev. Lett*. **102** (2009) 117602.

[7] C. Ederer, T. Harris, and R. Kovacik, Mechanism of ferroelectric instabilities in non-$d^0$ perovskites: $LaCrO_3$ versus $CaMnO_3$ *Phys. Rev. B* **83** (2011) 054110.

[8] S. Donga, J. M. Liub, S. W. Cheong, and Z. Rend, Multiferroic materials and magnetoelectric physics: symmetry, entanglement, excitation, and topology *Adv. Physics* **64** (2015) 519-626.

[9] W. Kleemann, Disordered Multiferroics *Solid State Phenomena* **189** (2012) 41-56.

[10] V. V. Shvartsman, S. Bedanta, P. Borisov, W. Kleemann, A. Tkach, and P. M. Vilarinho, $(Sr, Mn)TiO_3$: A Magnetoelectric Multiglass, *Phys. Rev. Lett*. **101** (2008) 165704.

[11] J. Kumar and A. M. Awasthi Quantum paraelectric glass state in $SrCu_3Ti_4O_{12}$, *Appl. Phys. Lett.* **104** (2014) 262905.

[12] D. Choudhury, P. Mandal, R. Mathieu, R. Mathieu, A. Hazarika, S. Rajan, A. Sundaresan, U. V. Waghmare, R. Knut, O. Karis, P. Nordblad, and D. D. Sarma Near-Room-Temperature Colossal Magnetodielectricity and Multiglass Properties in Partially Disordered $La_2NiMnO_6$ *Phys. Rev. Lett.* **108** (2012) 127201.

[13] R. Rawat, D. M. Phase, and R. J. Choudhary, Spin-phonon coupling in hexagonal $Sr_{0.6}Ba_{0.4}MnO_3$, *J. Magn. Magn. Mater.* **441** (2017) 398–403

[14] S. Chaturvedi, P. Shyam, R. Bag, M. M. Shirolkar, J. Kumar, H. Kaur, S. Singh, A. M. Awasthi, and S. Kulkarni, Nanosize effect: Enhanced compensation temperature and existence of magnetodielectric coupling in $SmFeO_3$, *Phys. Rev. B* **96** (2017) 024434.

[15] J. F. Scott and J. Gardner, Ferroelectrics, multiferroics and artifacts: Lozenge-shaped hysteresis and things that go bump in the night, *Materials Today* **21** (2018) 553.

[16] F. Schrettle, P. Lunkenheimer, J. Hemberger, V. Y. Ivanov, A. A. Mukhin, A. M. Balbashov, and A. Loidl Relaxations as Key to the Magnetocapacitive Effects in the Perovskite Manganites, *Phys. Rev. Lett.* **102** (2009) 207208.

[17] C. R. Serrao, A. K. Kundu, S. B. Krupanidhi, U. V. Waghmare, and C. N. R. Rao Biferroic $YCrO_3$, *Phys. Rev. B* **72** (2005) 220101(R).

[18] T. A. Tyson, T. Wu, K. H. Ahn, S. B. Kim and S. W. Cheong, Local spin-coupled distortions in multiferroic hexagonal $HoMnO_3$ *Phys. Rev. B* **81** (2010) 054101.

[19] S Hashemizadeh, A Biancoli, and D Damjanovic, Symmetry breaking in hexagonal and cubic polymorphs of $BaTiO_3$, *J. Appl. Phys*. **119** (2016) 094105.

[20] A Singh, V Pandey, R. K. Kotnala, and D. Pandey, Direct Evidence for Multiferroic Magnetoelectric Coupling in $0.9BiFeO_3$–$0.1BaTiO_3$, *Phys. Rev. Lett*. **101** (2008) 247602.

[21] R Raghunathan, J. P. Sutter, L. Ducasse, C. Desplanches, and S. Ramasesha, Microscopic model for high-spin versus low-spin ground state in $[Ni_2M(CN)_8]$ (M =$Mo^V$, $W^V$, $Nb^{IV}$) magnetic clusters, *Phys. Rev. B* **73** (2006) 104438.

[22] D. M. Nalecz, R. J. Radwanski & Z. Ropka, The crystal field effects in hexagonal 4H-$SrMnO3$, Phase Transit., 90 (2017), pp. 125-130.

[23] S. Sahoo, R. Raghunathan, S. Ramasesha, and D. Sen, Fully symmetrized valence-bond based technique for solving exchange Hamiltonians of molecular magnets, *Phys. Rev. B* **78** (2008) 054408.

[24] F. Moussa, M. Hennion, J. Rodriguez-Carvajal, and H. Moudden, L. Pinsard and A. Revcolevschi *Phys. Rev. B* 54 (1996), 15149.

[25] R. Søndenå, P. Ravindran, and S. Stølen, Electronic structure and magnetic properties of cubic and hexagonal $SrMnO_3$, *Phys. Rev. B* **74** (2006) 144102.

[26] S. Kamba, D. Nuzhnyy, M. Savinov, J. Šebek, and J. Petzelt, Infrared and terahertz studies of polar phonons and magnetodielectric effect in multiferroic BiFeO$_3$ ceramics, *Phys. Rev. B* **75** (2007) 024403.

[27] S. Pandey, J. Kumar, and A. M. Awasthi, Magnetodielectric behaviour in La$_{0.53}$Ca$_{0.47}$MnO$_3$, *J. Phys. D: Appl. Phys.* **47** (2014) 435303.

[28] G. Catalan, Magnetocapacitance without magnetoelectric coupling, Appl. Phys. Lett. **88** (2006) 102902.